\documentclass[11pt,twoside]{article}
\usepackage{asp2004}
\usepackage{psfig}
\usepackage{epsf}
\usepackage{graphics}
\usepackage{lscape}
\begin{document}
\title{Modeling the Circumstellar disk of $\zeta$ Tauri}
\author{Alex C. Carciofi,}
\affil{Instituto de Astronomia, Geof\'\i sica e Ci\^encias Atmosf\'ericas / USP,
Rua do Mat\~ao 1226, Cidade Universit\'aria,
05508-900 S\~ao Paulo SP - Brazil}
\author{Jon E. Bjorkman and Karen S. Bjorkman}
\affil{Department of Physics and Astronomy, MS 111
The University of Toledo, 2801 West Bancroft Street,
 Toledo, Ohio 43606 
}

\begin{abstract}
We present a model for the disk of the classical Be star $\zeta$ Tauri.
The model consists of a Keplerian rotating disk with a power-law surface density and a vertical density distribution that follows from the balance between the thermal gas pressure and the z-component of the stellar gravitation. The opening angle of such a disk is not a fixed value but increases with the distance to the star (flared disk).
We use a Monte Carlo code that solves simultaneously the thermal equilibrium, the statistical equilibrium and the hydrostatic equilibrium equations. The result is a self-consistent solution for the electron temperature, level populations and gas density distribution.
It follows from this model that the only free parameter is the disk density. All other parameters (e.g. shape of the disk surface) are self-consistently calculated given the disk density. In this paper we present a successful fit for $\zeta$ Tauri data using this model.
\end{abstract}
\thispagestyle{plain}

\section{Stellar Parameters and Observations}

Following \citet{woo97},  we adopt for $\zeta$ Tauri a luminosity class of IV and a spectral type of $\rm B2.9 \pm 0.4$. 
This corresponds to an effective temperature of $19,000 \rm K$ and a stellar radius of $5.6 R_{\sun}$.

We use spectroscopic and spectropolarimetric data from the half-wave plate spectropolarimeter (HPOL) on the Universty of Wisconsin's PBO. For details on data acquisition and reduction see \citet{woo97}. In addition to the optical data we use archival IRAS fluxes at 12, 25 and 60 $\mu \rm m$.

\section{Hydrogen Disk Properties}

We adopt a viscous decretion alpha disk composed of pure Hydrogen, which rotates with Keplerian velocity and has a steady-state power-law surface density. The vertical density structure of the disk is obtained from the solution of the hydrostatic equilibrium (HSEQ) equations, and is self-consistently determined in conjunction with the disk temperature structure and level populations.

It is interesting to note that the only free parameter  (aside from the stellar parameters and the outer radius of the disk) is the disk density scale.

\section{Model Results}

Our best fit model corresponds to a density at the base of the star of $\rho_0=7.5 \times 10^{-11} \rm g/cm^3$ and an inclination angle of $79 \deg$. In Figure 1 we display the disk temperature structure. It is interesting to note the drop of the temperature in the equator and the fact that the disk becomes nearly isothermal for large radii.
In Figure 2 we show our results for the optical SED, IR SED and polarization.

\begin{figure}[]
\plotone{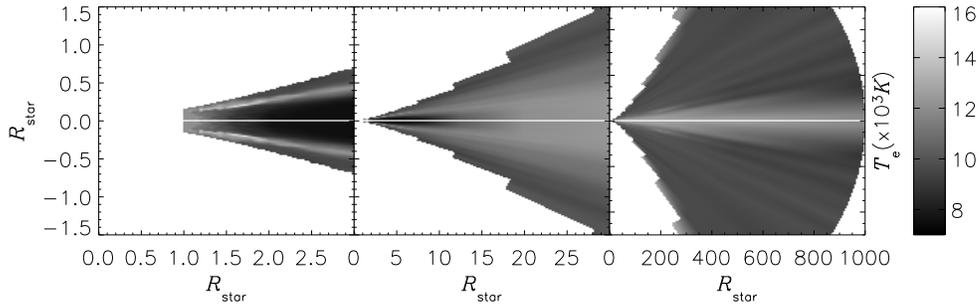}
\caption{Temperature structure of the HSEQ solution of a Keplerian disk.}
\end{figure}

\begin{figure}[]
\plotone{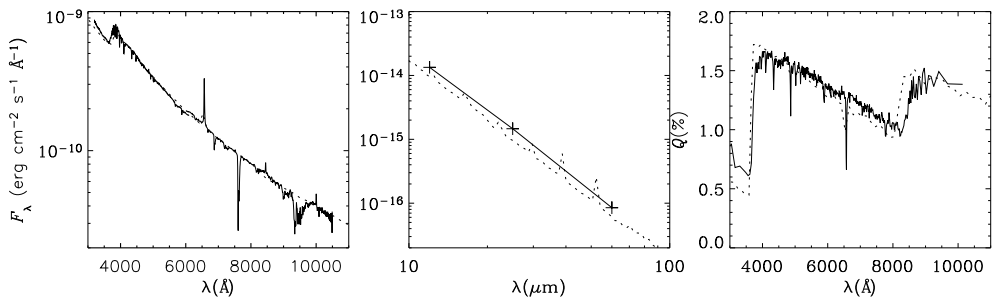}
\caption{Observations (solid lines) and model results (dashed lines) for $\zeta$ Tauri. Left and Center: flux. Right: linear polarization.}
\end{figure}

\section{Discussion}

Using our Keplerian disk model we were able to obtain a very reasonable fit for both the flux and polarization. Considering that our model is essentially a one-parameter model, we believe it represents a strong evidence in support for the existence of Keplerian or near Keplerian disks around classical Be stars.

The results we present are still preliminary. Our next step will be to include the available UV polarization data in our model, as well as Hydrogen line profiles.

\acknowledgements{This work has been supported by NSF Grant AST9819928 to the University of Toledo.}

\end{document}